\documentclass[10pt,letterpaper]{article}
\usepackage{opex3}

\newcommand{\Mit}{\mathrm}

\newcommand{\yb}{$\mbox{Yb}^+\:$}

\begin{document}

\title{Mode-locked picosecond pulse generation from an octave-spanning supercontinuum}

\author{D. Kielpinski and M.G. Pullen}

\address{Centre for Quantum Dynamics, Griffith University, Nathan QLD 4111, Australia}

\author{J. Canning and M. Stevenson}

\address{Interdisciplinary Photonics Laboratories (iPL), School of Chemistry, Madsen Building F09, University of Sydney, Sydney, NSW, 2006, Australia}

\author{P.S. Westbrook and K.S. Feder}

\address{OFS Laboratories, Somerset, New Jersey 08873, USA}

\email{d.kielpinski@griffith.edu.au} 

\begin{abstract} We generate mode-locked picosecond pulses near 1110 nm by spectrally slicing and reamplifying an octave-spanning supercontinuum source pumped at 1550 nm. The 1110 nm pulses are near transform-limited, with 1.7 ps duration over their 1.2 nm bandwidth, and exhibit high interpulse coherence. Both the supercontinuum source and the pulse synthesis system are implemented completely in fiber. The versatile source construction suggests that pulse synthesis from sliced supercontinuum may be a useful technique across the 1000 - 2000 nm wavelength range. \end{abstract}



\section{Introduction}

The advent of optical frequency counting techniques using octave-spanning supercontinuum (SC) generation \cite{Hall-frequency-metrology-rev, Udem-Hansch-optical-metrology-review} suggests that similar technology can prove useful in the synthesis of optical radiation. Much useful work has been done to show the possibilities of optical frequency synthesis by phase-locking a stable single-frequency laser to an optical frequency counter. Such techniques can achieve ultimate frequency resolution of a part in $10^{19}$ \cite{Ma-Diddams-optical-frequency-synthesis} with broad and fast tunability over wide frequency ranges \cite{Jones-Hall-arbitrary-wavelength-fs, Schibli-Matsumoto-optical-frequency-synthesis}. More direct approaches obtain highly coherent light by amplifying a single comb line \cite{Vainio-Nyholm-comb-injected-diode} or a band of comb lines \cite{Cruz-Ye-diode-amplified-comb}, or by injection-locking an external laser oscillator with a single comb line \cite{Moon-Park-comb-injected-diode}. \\

On the time-domain side, spectral slicing of relatively narrowband ($< 200$ nm) supercontinua has been used to generate picosecond pulse trains in hundreds of wavelength channels at once \cite{Morioka-Saruwatari-first-spectral-slicing, Smirnov-Turitsyn-telecom-optical-synthesis-rev} and these pulses are known to have coherence times greater than the pulse duration \cite{Boyraz-Jalali-10Gb-supercontinuum-pulse}. Raman self-frequency shifting also allows the generation of ps pulses at various wavelengths, and again the coherence time is known to be greater than the pulse duration \cite{Price-Richardson-tunable-raman-source}. However, the optical frequency synthesis applications enabled by octave-spanning SC require pulse-to-pulse phase coherence over times much longer than the repetition period of the pulse train to ensure that the mode-locked pulse train gives rise to a stable optical frequency comb. Our work is the first to investigate the temporal coherence of sliced octave-spanning supercontinuum over timescales greater than the pulse duration. \\

Here we describe the generation of mode-locked picosecond pulses from an octave-spanning supercontinuum (SC) source pumped at 1550 nm. The 1110 nm wavelength is of particular interest for our experiments on quantum computing with trapped \yb ions, but the architecture demonstrated here can be adapted to a wide variety of wavelengths in the 1000-2000 nm range of our SC source. The characteristics of our 1110 nm source are similar to those of a picosecond mode-locked Yb fiber laser in all respects but wavelength (see, e.g., refs. \cite{Porta-Traynor-ps-ML-Yb-fiber, Gomes-Okhotnikov-Yb-SESAM-laser}). The picosecond output pulses are chirped but can be compressed nearly to the transform limit, while the pulse energy of $\sim 100$ pJ and the 40 MHz repetition rate are similar to those of a typical mode-locked fiber laser. To our knowledge, no such laser has been reported at 1110 nm, showing that our architecture can expand the wavelength range of mode-locked fiber sources. Our results indicate that the 1110 nm light inherits the coherence properties of the SC source. In the future, absolute frequency stabilization of our SC source will enable us to coherently synthesize picosecond pulses at desired wavelengths, representing a major step toward an arbitrary optical field synthesizer. \\

\section{Octave-spanning supercontinuum}

\begin{figure}[htbp]
\centering\includegraphics[width=12cm]{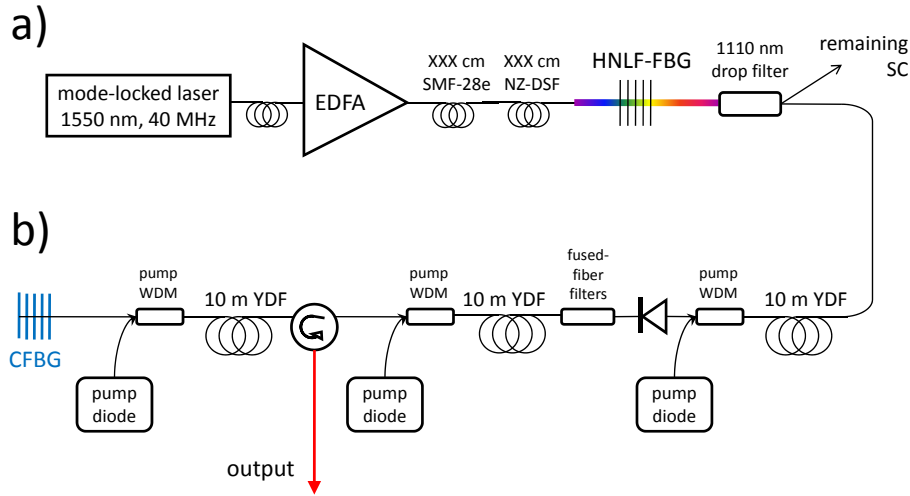}
\caption{Schematic of the mode-locked 1110 nm source. Supercontinuum generation (a) provides a spectrally sliced seed for reamplification (b).}
\label{schem}
\end{figure}

\begin{figure}[htbp]
\centering\includegraphics[width=7cm]{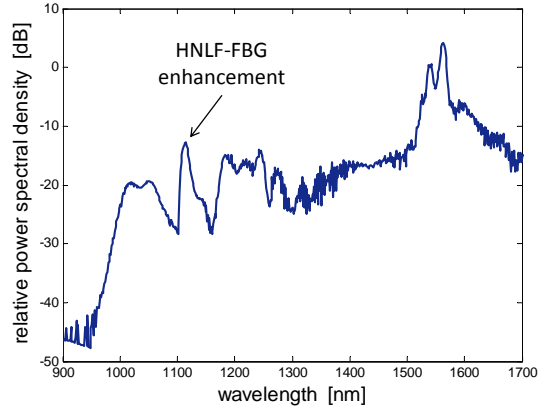}
\caption{Spectrum of supercontinuum generated by the 1550 nm pump showing coverage of the 1000-1700 nm wavelength range.}
\label{scspec}
\end{figure}

We generate supercontinuum (SC) in a highly nonlinear fiber Bragg grating pumped by femtosecond pulses at 1550 nm (Fig. \ref{schem}(a)). The 1550 nm signal originates in a mode-locked fiber laser (Precision Photonics) with 40 MHz repetition rate and is amplified to $\sim 120$ mW in an Er-doped fiber amplifier (IPG Photonics). The lengths of single-mode fiber (SMF) at the input and output of the power amplifier are optimized to give a minimum pulse duration of 90 fs at the amplifier output. Another 20 cm length of dispersion-shifted fiber (NZ-DSF) compresses the amplified pulse further, to 55 fs, though the soliton compression leaves behind a pedestal of several ps duration containing about 50\% of the pulse energy. The amplified pulses are injected into a highly nonlinear fiber Bragg grating (HNLF-FBG) with bandgap near 1108 nm. A Bragg grating was written into the HNLF using a 248 nm excimer beam and a phase mask.   The 6 mm UV beam was scanned uniformly for 27 mm.  The resulting grating had a core mode reflection bandwidth extending from 1104.4nm to 1108.4nm. The HNLF extends another 30 cm past the FBG region. Fig. \ref{scspec} shows the SC spectrum in the 900 - 1700 nm wavelength range, as measured with an optical spectrum analyzer (OSA). While the OSA range is limited to wavelengths below 1700 nm, further measurements with an IR monochromator indicate that the SC plateau extends to $>2000$ nm. Splice losses and linear loss in the FBG reduce the SC average power to 40 mW at the HNLF output. The SC power spectral density is enhanced near 1110 nm, slightly to the red of the FBG bandgap \cite{Westbrook-Feder-grating-SC-enhancement}. The SC source maintains high spectral brightness ($> 10\: \mu\mbox{W}/\mbox{nm}$) over nearly the entire 1000 - 2000 wavelength range. This spectral brightness is sufficient for generation of mode-locked picosecond pulses at mW power levels over substantial portions of the SC spectrum, as exemplified in the present work at 1110 nm.  \\

\vspace{1cm}

\section{Spectral slicing and reamplification}

The SC source is initially spectrally sliced by custom fused-fiber couplers to isolate the 1000 - 1300 nm band. This part of the SC serves as the input to a home-built ytterbium-doped fiber amplifier (YDFA) system (Fig. \ref{schem}(b)) that preferentially amplifies input light near 1110 nm. While the gain peak of short lengths of ytterbium-doped fiber is near 1030 nm, excited-state absorption at this wavelength is also relatively high, so that longer lengths of more highly doped fiber exhibit gain peaks at relatively longer wavelengths. Our YDFA stages use 10-20 meter lengths of fiber with absorption coefficient of 1200 dB/m at 980 nm and exhibit gain peaks in the 1080 - 1110 nm range. Each stage is backward-pumped with up to 400 mW of 976 nm diode laser light. The first YDFA stage amplifies a relatively broad SC band from 1060-1120 nm. An optical isolator separates the first two YDFA stages, minimizing parasitic lasing. Custom fused-fiber couplers filter out wavelengths below 1085 nm at the input to the second YDFA stage. A fiber circulator passes the second-stage output through to the third YDFA stage. After a first pass through this stage, the amplified SC encounters a chirped fiber Bragg grating (CFBG). This CFBG ($L \sim 4$ mm, $\Delta \lambda_\Mit{chirp} \sim 1.1$ nm, $R \sim 7.4$ dB) was fabricated using a computer controlled, scanning free space Sagnac interferometer configuration where the 244nm output of a frequency doubled Ar+ laser is split mostly into the two first order diffraction orders through a chirped phase mask (pitch 10678.5nm) designed for 1554nm (chirp $\sim 3 \mbox{nm/cm}$). By tilting the interferometer mirrors, the translated chirped interferogram has a reduced pitch which is imaged onto the custom made germanosilicate optical fibre (cutoff wavelength $<700$ nm) by the Sagnac interferometer. The final Bragg wavelength is centered at $\sim 1108.5$ nm after annealing. The CFBG reflection bandwidth is approximately 1 nm (FWHM) around 1108.5 nm, which serves to filter out the parasitically amplified wavelengths below 1105 nm. The light reflected from the CFBG passes again through the third YDFA stage, being further amplified on the way, and emerges at the output port of the circulator. This narrowband double-passing of the final YDFA stage amplifies the 1110 nm signal by an additional 4 dB, while reducing the amplified SC background by 6 dB, leading to an overall improvement of 10 dB in spectral selectivity around 1110 nm. \\

\begin{figure}[htbp]
\centering\includegraphics[width=7cm]{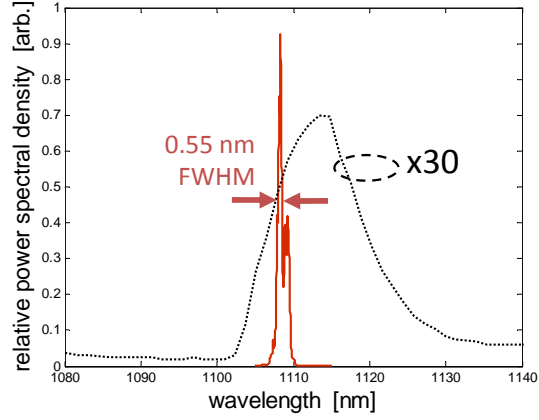}
\caption{SC input (dotted line) and YDFA output (solid line) power spectral density near 1110 nm. The input spectrum is multiplied by a factor of approximately 30 for ease of comparison.}
\label{ybspec}
\end{figure}

Figure \ref{ybspec} shows the optical power spectra at the YDFA system's input and output. While the input spectrum extends far outside the displayed limits, the wavelength range amplified by the YDFA is set by the 1 nm bandwidth of the CFBG. From the measured $\sim 200\: \mu$W power contained in the 5 nm wide SC enhancement peak, we see that the in-band power input to the YDFA is only $40\: \mu$W. Total output powers of up to 5.5 mW are achieved, with a net amplification of 21 dB. \\

\section{Pulse properties and coherence}

\begin{figure}[htbp]
\centering\includegraphics[width=7cm]{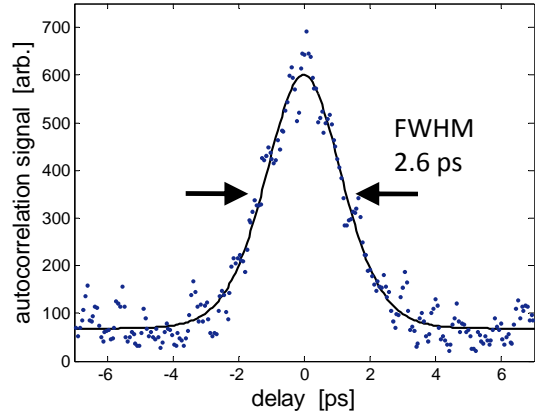}
\caption{Autocorrelation trace of the compressed 1110 nm pulses. Points: data; line: sech${}^2$ fit. The autocorrelation signal is offset from zero for ease of viewing.}
\label{actrace}
\end{figure}

The 1110 nm pulses are initially highly chirped but can be compressed to nearly the transform limit. Measured at the output of the YDFA, the pulses have full-width at half-maximum (FWHM) duration of $\sim 6$ ps. (We assume sech${}^2$ pulse shape throughout.) We use a free-space grating compressor to compensate the normal group-velocity dispersion (GVD) accumulated in the YDFA. Figure \ref{actrace} shows an autocorrelation trace for optimal pulse compression. The corresponding pulse duration is $1.7 \pm 0.1$ ps FWHM. We must be cautious in estimating the time-bandwidth product because of the asymmetric source spectrum (Fig. \ref{ybspec}); the spectral FWHM is only 0.7 nm, yielding a time-bandwidth product smaller than the transform limit. Numerically predicting the transform-limited pulse duration for the measured output spectrum, we find that the 1110 nm pulses are within a factor of 1.6 of the transform limit. No pulse pedestal is visible within the 50 ps scan range of the autocorrelator, showing that the pulses are of high temporal quality. \\

The 1110 nm pulses are ``mode-locked'' only if the temporal coherence of the electric field is maintained from one pulse to the next. Here we rely on the SC generation process to transfer the coherence of the 1550 nm pump pulses to the derived 1110 nm pulses. Unless the pump pulse duration is short ($<100$ fs) and the soliton number in the HNLF is low ($<15$), the coherence of the SC process can easily be lost through fundamental mechanisms such as photon shot noise and spontaneous emission \cite{Dudley-Coen-fiber-SC-rev}. Our pump source has a 55 fs pulse duration and an estimated soliton number of $\sim 9$, so high SC coherence is expected. \\

We measure the temporal coherence between successive 1110 nm pulses with an unbalanced Mach-Zehnder interferometer implemented in fiber. A piezoelectric stretcher in one arm of the interferometer sweeps the path length difference $\Delta L$, and we observe interference fringes by photodetection when $\Delta L$ is close to the distance $L_\Mit{rep}=c/\nu_\Mit{rep}$ between successive pulses. The fringe visibility is limited by the differential dispersion of the interferometer arms. We use 500 cm of Corning SMF28e fiber, with dispersion of $-22 \mbox{ps}/(\mbox{nm km})$ \cite{CorningSMF28e}, to implement the distance $L_\Mit{rep}$. Numerically simulating the dispersive effect of the fiber on the measured YDFA spectrum, we obtain a maximum fringe visibility of approximately 55\%, so the measured visibility of $49 \pm 1$\% indicates a high degree of interpulse coherence. \\

\begin{figure}[htbp]
\centering\includegraphics[width=7cm]{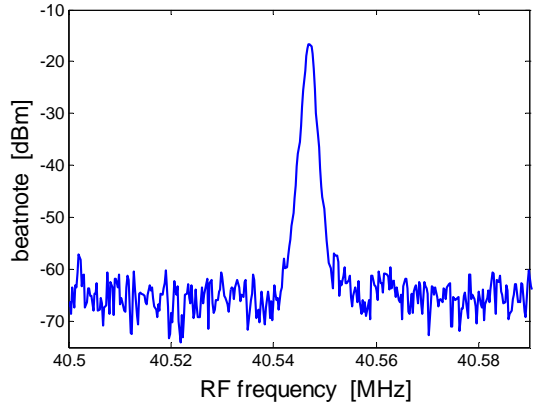}
\caption{RF spectrum of the 1110 nm source. Resolution bandwidth: 1 kHz.}
\label{rfspec}
\end{figure}

The RF spectrum produced by photodetection of a mode-locked laser source contains a strong, narrow line at the laser repetition rate $\nu_{rep}$, indicating low timing jitter between the envelopes of successive pulses. Figure \ref{rfspec} shows the RF spectrum of the 1110 nm source as measured by a fast photodiode. The RF spectrum is indistinguishable from that of the original 1550 nm oscillator at our minimum resolution bandwidth of 1 kHz. \\

\vspace{2.5cm}

\section{Outlook}

We have generated mode-locked picosecond pulses at 1110 nm by spectral slicing and reamplification of an octave-spanning supercontinuum source. The resulting light source is similar to that of a standard mode-locked fiber oscillator, except that such oscillators have not, to our knowledge, been demonstrated at this wavelength. Much of the complexity of our amplifier design arises from the low gain at our target wavelength of 1110 nm, exacerbated by the relatively low quality of fiber components at this unusual wavelength. Hence our experiment demonstrates the feasibility of all-fiber mode-locked ps pulse generation right across the ytterbium-doped fiber gain bandwidth of 1010 - 1150 nm. Pressure tuning of a suitable wavelength-selecting CFBG would enable dynamic tuning of the ps pulse wavelength over $>100$ nm \cite{Mokhtar-Ibsen-pressure-FBG-tuning}. Amplification of a spectrally broader SC seed should permit generation of femtosecond pulses, as long as the nonlinear and dispersive pulse propagation in the amplifier is properly taken into account. \\

More broadly, fiber and semiconductor amplifiers can be used in place of our YDFA over most of the 1000 - 2000 nm wavelength range. Taking the 1110 nm source bandwidth of $\sim 1.5$ nm as typical, we see that several hundred wavelength channels of mode-locked ps pulses can be simultaneously derived from a single SC source, as shown in other work at lower SC bandwidth. As we have shown, each channel can be made to inherit its phase behavior from the SC source, passively phase-locking all channels together. Stabilizing the SC source in absolute frequency, combined with independent modulation of each channel, would allow phase-coherent synthesis of high-resolution optical field waveforms.\\

\section{Acknowledgments}

This work was supported by the US Air Force Office of Scientific Research under contract FA4869-06-1-0045, by DIISR ISL Grant CG130013, and by the Australian Research Council (ARC) under DP0773354 (Kielpinski), DP0770692 (Canning), and Prof. Howard Wiseman's Federation Fellowship FF0458313. We thank Go{\"e}ry Genty, {\"O}mer Ilday, and Nathan Newbury for helpful discussions.

\end{document}